\pdfoutput=1
\documentclass[aps,prb,twocolumn,superscriptaddress,showpacs, floatfix]{revtex4-1}
\usepackage{amssymb}
\usepackage{graphicx}
\usepackage{amsmath}
\usepackage{amsfonts}
\usepackage{amssymb}%
\usepackage{gensymb}
\usepackage[utf8x]{inputenc}

\begin{document}
\title{Hybrid-functional electronic structure of multilayer graphene.}

\author{Marco Campetella}
\email[]{marco.campetella@upmc.fr}
\author{Nguyen Minh Nguyen}
\author{Jacopo Baima}
\affiliation{Sorbonne Universit\'e, CNRS, Institut des Nanosciences de Paris, UMR7588, F-75252, Paris, France}
\author{Lorenzo Maschio}
 \affiliation{Dipartimento di Chimica and Centre of Excellence NIS (Nanostructured Interfaces and Surfaces), Universit\`{a} di Torino, via P. Giuria 5, I-10125 Turin, Italy}
\author{Francesco Mauri}
 \affiliation{Dipartimento di Fisica, Universit\`{a} di Roma La Sapienza,
Piazzale Aldo Moro 5, I-00185 Roma, Italy}
\author{Matteo Calandra}
\email[]{matteo.calandra@upmc.fr}
\affiliation{Sorbonne Universit\'e, CNRS, Institut des Nanosciences de Paris, UMR7588, F-75252, Paris, France}

\begin{abstract}
Multilayer graphene with rhombohedral and Bernal stacking are supposed to be metallic, as predicted by density functional theory calculations using semi-local functionals. However recent angular resolved photoemission and transport data have questioned this point of view. In particular, rhombohedral flakes
 are suggested to be magnetic insulators, a view supported also by magnetic hybrid functional calculations. Bernal flakes composed of an even number of layers are insulating (for N $\le 6$), while those composed of an odd number of layers are pseudogapped (for N $\le 7$). 
Here, by systematically benchmarking with plane waves codes, we develop very accurate all-electron Gaussian basis sets for graphene multilayers leading to a precise
description of the single-particle electronic structure in the 100 meV energy range from the Fermi level. We find that, in agreement with our previous calculations, rhombohedral stacked multilayer graphene are gapped
for ($N\ge3$) and magnetic. However, the valence band curvature and the details of the electronic structure depend crucially on the basis set. Only substantially extended basis sets are able to correctly reproduce the effective mass of the valence band top at the K point, while the popular POB-TZVP basis set leads to a severe overestimation. 
In the case of Bernal stacking, we show that exact exchange gaps the flakes composed by four layers and opens  pseudogaps for N $=3,6,7,8$. However, the gap or pseudogap size and its behaviour as a function of thickness are not compatible with experimental data. Moreover, hybrid functionals lead to a metallic solution for $5$ layers and a magnetic ground state for $5$, $6$ and $8$ layers. Magnetism is very weak with practically no effect on the electronic structure and the magnetic moments are mostly concentrated in the central layers. Our hybrid functional calculations on trilayer Bernal graphene multilayers are in excellent agreement with non-magnetic GW calculations. For thicker multilayers, our calculations are a benchmark for manybody theoretical modeling of the low energy electronic structure.

\end{abstract}

\maketitle

\section{Introduction}

\begin{figure}[!h]
\includegraphics[width=1.05\linewidth]{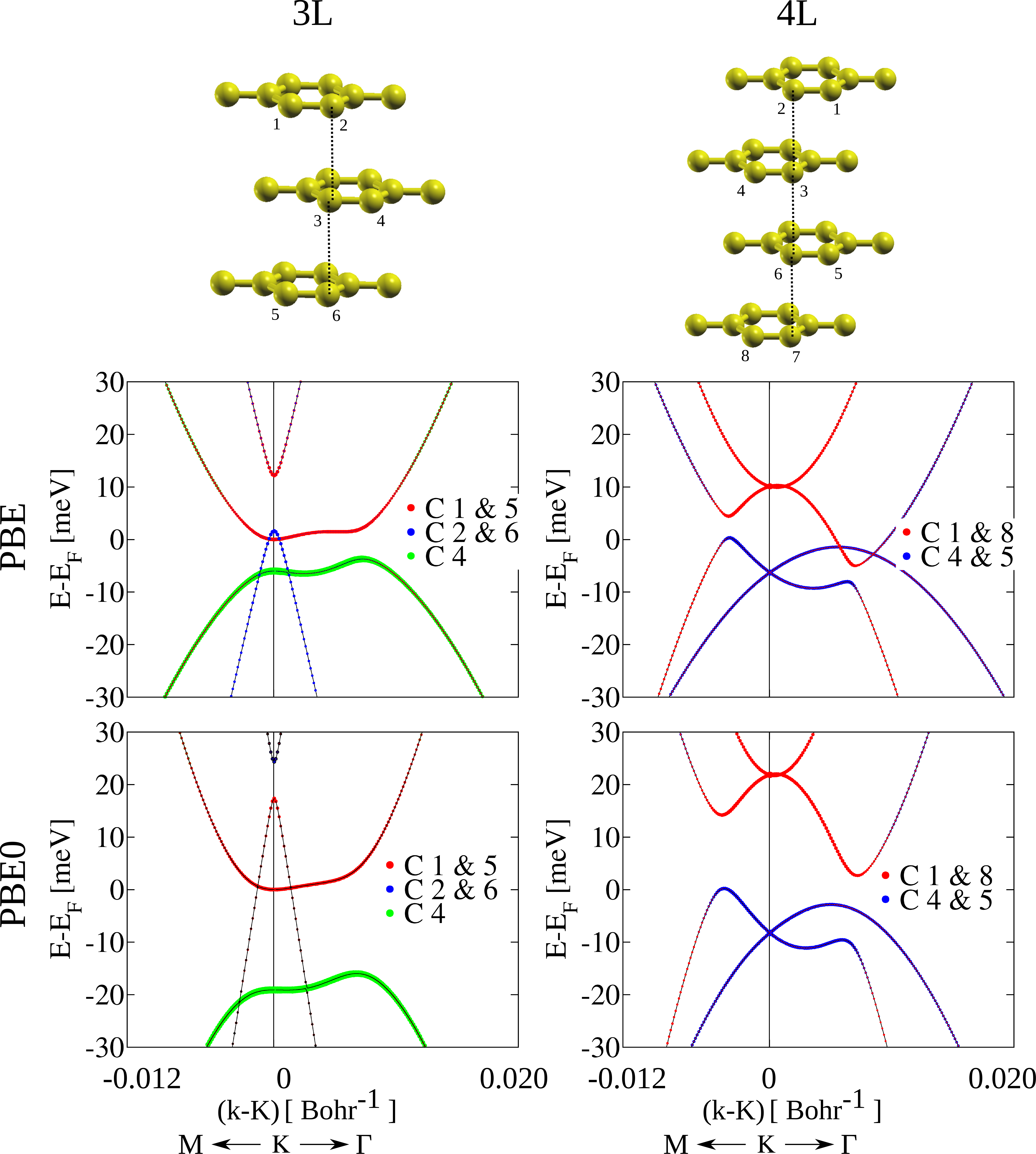}
\caption{Top panels: Atom numbering scheme for three and four layers Bernal graphene. 
Bottom panels: PBE and PBE0 electronic  structure of trilayer (left panels) and four-layer (right panels) Bernal graphene. The color is proportional to the p$_z$ orbital character of the Carbon atoms. The electronic bands are plotted around the \textbf{K} point, that has been chosen as origin. }
\label{fig:bandepbe0}
\end{figure}

The electronic structure of multilayer graphene has been calculated with a variety of techniques
such as tight-binding with parameters fitted on experiments \cite{koshino2013stacking,partoens2006graphene,KoshinoSSC,PhysRevB.80.165409}, 
the effective mass approximation or low energy expansions\cite{koshino2007orbital,min2008electronic}, density functional theory
with semilocal functionals (typically LDA or PBE) \cite{latil2006charge,aoki2007dependence}
and non-magnetic RPA  \cite{PhysRevB.88.085439} (bilayer graphene) and GW approximations \cite{PhysRevB.88.085439,PhysRevB.89.035431} (bilayer and trilayer graphene). 
The view emerging from all these calculations is that all systems are metallic/semimetallic 
(i.e. they do not have a gap). In more details, close to the Fermi level, at  the special point K of the Brillouin zone, rhombohedral stacked multilayers display a flat surface state. Bernal stacked multilayers show
 metallic massive bands for even $N$, while a Dirac cone coexists
with massive bands for odd N (see Fig. \ref{fig:bandepbe0} for the $N=3,4$ case). 

Several experiments contradict this view. Transport measurements on unsupported rhombohedral flakes composed of three and four layers\cite{Lautrilayer,Lauquadrilayer} show the presence of an insulating state with gaps larger than 40 meV. Recent magnetotransport experiments
on unsupported ABC trilayers display large and field-effect tunable magnetoconductance histheresys, suggestive of a magnetic state 
\cite{LauCalandra}. At larger n-doping, the magnetic state is predicted to melt in an half-metallic ground state \cite{RMG_HM}. Two layers of bilayer graphene  twisted by tiny angles have been shown to form uniform
four layer ABCA graphene regions with a 9.5 meV gap as measured in STM and attributed to manybody effects\cite{Angel}. 

Samples of rhombohedral stacked graphene with thickness up to 50 layers were recently isolated
\cite{Henni_NL, RMGhenck,MischenkoABC}. The rhombohedral stacking was identified via Raman spectroscopy \cite{Henni_NL,Torche}  and Landau level measurements \cite{Henni_NL, MischenkoABC}.

ARPES data on 14 layers samples were found to be
consistent with the occurrence of a magnetic state\cite{RMGhenck}, by comparison of the valence band effective mass at the K point with hybrid functional calculations for magnetic and non-magnetic solutions\cite{pamuk2017magnetic,RMGhenck}. The curvature of the top of the valence bands at K was found to be much larger in the magnetic case than in the non magnetic case.

The situation is similar for Bernal stacking, as several measurements suggest the occurrence of a gapped state  on suspended samples \cite{freitag2012spontaneously,nam2016,nam2018family, grushina2015insulating}. Among them, a very recent paper \cite{nam2018family},
show that all N-layer Bernal suspended graphene flakes with $2\le N\le 7$ are insulating or pseudogapped. Specifically, the resistance at the charge neutrality point of suspended flakes with $N=2,4,6$ is in the range $5\times10^3$ to $5\times10^5$ k$\Omega$ at $T=0.25$K, monotonically increasing with thickness. In the odd number of layer case,
the resistance at charge neutrality is smaller and of the order of $25$ to $55$ k$\Omega$ at $T=0.25 K$ , monotonically decreasing with thickness. The behaviour of
flakes with an odd number of layers is more suggestive of a pseudogapped phase than that of a completely insulating state. For even $N$
the gap measured via transport increases with thickness and ranges between $1$ and $13$ meV, substantially smaller than for the case of ABC multilayers. 

\begin{figure}[!h]
\includegraphics[width=0.5\linewidth]{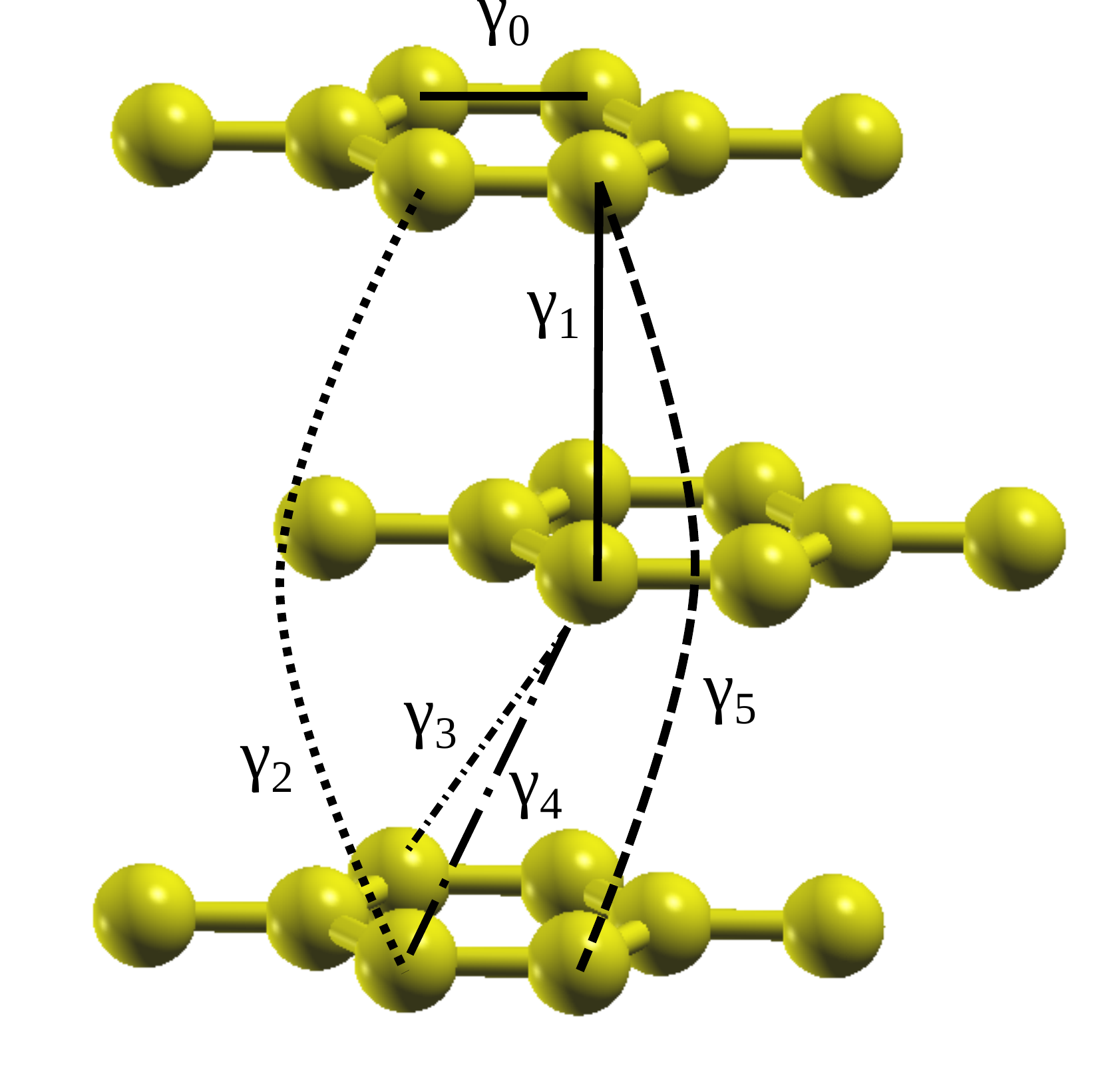}
\caption{Tight-binding hopping parameters for multilayer graphene with Bernal stacking.}
\label{fig:hopping}
\end{figure}

In Bernal graphene multilayers, the massive bands close to the Fermi level have small Fermi velocities and large effective masses. Given the small kinetic energy of the
electrons in these bands,  a gap could open due to electron-electron interaction  effects not included in semilocal functionals, as it happens for ABC graphene multilayers\cite{pamuk2017magnetic}. Carbon based systems hosting correlated states are not so uncommon, as this is actually proposed to 
happen in graphene multilayers with rhombohedral (ABC) stacking\cite{Henni_NL,pamuk2017magnetic,RMGhenck,Lautrilayer,Lauquadrilayer}, in twisted bilayer graphene\cite{cao2018unconventional,cao2018correlated} or in diamond(111)\cite{Betul_Diamond}. In all these cases, the correlated
state is proposed to emerge from flatbands. From the theoretical point of view, a self-consistent tight-binding calculation \cite{Koshino_Exchange} with empirical inclusion of Hartree and exchange terms on
Bernal multilayers suggests that the  exchange interaction substantially modifies the electronic structure, via the renormalization of the
$\gamma_2$ hopping term (see Fig. \ref{fig:hopping} for the hopping processes ). However, the possible occurrence of magnetism was not studied in this
work, and the empirical form of the electron-electron interaction calls for more accurate calculations.

A first step towards the understanding of the electron-electron interaction effects in multilayer graphene is, then, the determination of the role of the
exchange interaction. This can be quantitatively evaluated at the mean field level by using hybrid functionals including a certain percentage of Hartree-Fock exchange. In single layer graphene the renormalization of the Fermi velocity is well captured by hybrid functionals\cite{stauber2017interacting}. The difficulty is, however, that hybrid functional calculations are computationally demanding, particularly in the case of multilayer graphene with Bernal stacking as the Fermi surface is very narrow (linear dimension of $\approx 0.01$ \AA$^{-1}$) and the Brillouin zone sampling becomes soon prohibitive, particularly, if plane wave codes are used. This difficulty has hindered, up to now, calculations beyond semilocal functionals in these systems. 

Here, by developing a very accurate basis set taylored for multilayer graphene, we perform
all-electron electronic structure calculations with the inclusion of exact exchange and ultradense  Brillouin zone sampling (up to $1200\times 1200$ for
the self-consistent calculation and up to $12000\times12000$ for the density of states). We carry out an in-depth analysis of the electronic
structure and compare our results with experimental data. We consider (6 and 14 layers) both the case of thick ABC graphene samples and Bernal stacked flakes up to $7$ layers.

The paper is organised as follows. In section \ref{sec:computationaldet} we
describe the technical details of the calculation. In
sec. \ref{sec:results} we analyze the possible stabilization of magnetic state in multilayer Bernal graphene. Finally, we discuss the electronic structure of these systems.  

\section{Computational Details\label{sec:computationaldet}}

\subsection{Simulation parameters}

All electron electronic structure calculations were performed using the CRYSTAL code\cite{dovesi2014crystal14}. The PBE and the PBE0\cite{adamo1999toward} hybrid functionals  have been used for DFT calculations with an \textit{ad hoc} optimized def2-TZVP Gaussian-type basis sets\cite{weigend2005balanced} for the C atoms labeled as TZVP+ (see section (\ref{sec:QZVP+}) for more details) for Bernal graphene and a QZVP adapted basis set for rhombohedral graphene. The Gaussian exponents and coefficients are reported in the supporting information (SI). In order to avoid numerical instabilities due to an ill-conditioned overlap matrix (basis set near-linear dependency), we removed the eigenvectors belonging to the smallest eigenvalues with a thereshold of 10$^{-5}$. The band flatness and the extreme localization of the low-energy states around the special point \textbf{K} require an ultra-dense sampling with an electronic k mesh of $1200\times1200$. 
The real space integration tolerances was set to 11 11 11 15 40 (in order to use such extended basis set) and with an energy tolerance of 10$^{-11}$ Ha for the total energy convergence. The density of states (DOS) is obtained with a Gaussian
smearing of 0.00001 Ha. The grid points on which the DOS has been calculated is characterized by a square grid centered on the
\textbf{K} point. The square edge extension is $3/10$ of the length of the reciprocal space basis
vector, and the total number of k points used is $641601$. It is equivalent to a k mesh of $12000\times12000$ on the entire BZ. Such dense grid was needed to converge the DOS. In the case of magnetic calculations, we fix the magnetic state in the first iteration of the self-consistent cycle, and then we release the constraint. We worked in fixed geometry and we have chosen an in-plane lattice parameter of a = 2.461 \AA\ and an inter-plane distance of 3.347 \AA. The distance between two adjacent periodic images is 13.38 \AA\ along the z direction. Moreover for the trilayer and quadrilayer bernal graphene, density functional theory calculations with the PBE functional are performed using the Quantum-Espresso\cite{giannozzi2009quantum,giannozzi2017advanced} code as well, in order to verify the consistency between pseudopotential
and Gaussian basis set. In this case, for carbon
we use norm conserving and PAW pseudopotentials. We use an energy cutoff up to 65 Ry for all calculations. For the exchange correlation energy we adopt the generalized gradient approximation (GGA). The charge density integration over the Brillouin Zone (BZ) is performed using an uniform $512 \times 512$ Monkhorst and Pack grid\cite{monkhorst1976special}. The geometry of the systems is the same used for CRYSTAL.
In PBE, the electronic structures are consistent within $2-3$ meV and do not present qualitative differences. 
It is, however, difficult to infer if the residual differences come from the fact of using a Gaussian basis set instead of plane waves or
 if it is due to the use of an all-electron calculation against a pseudopotential one. 

\subsection{Accuracy of Gaussian basis set \label{sec:QZVP+}}

\begin{figure*}[!h]
%\begin{minipage}[c]{0.2\linewidth}
\includegraphics[scale=0.15]%angle=270]
{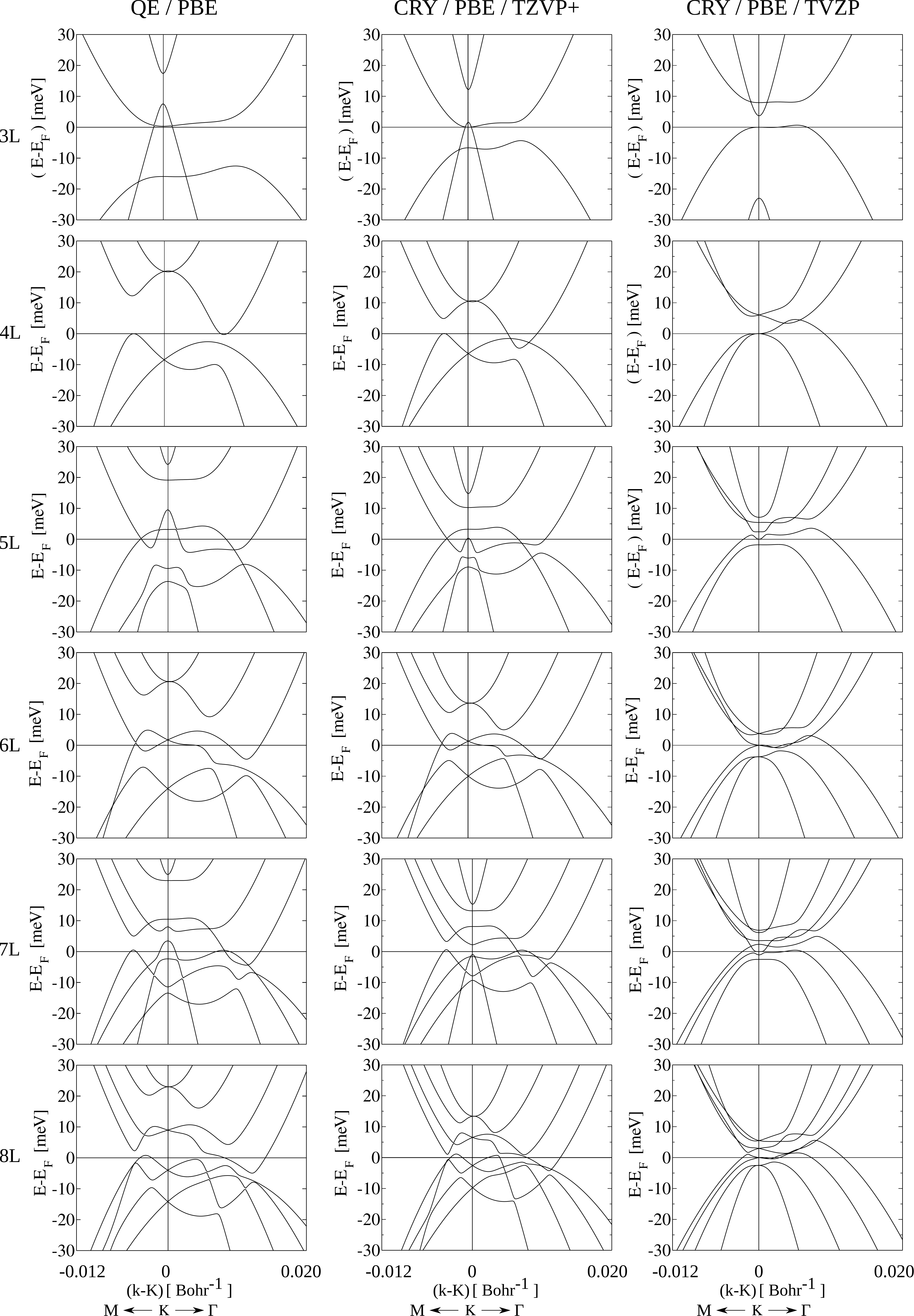}
\caption{PBE band structure of all  studied multilayer systems with Bernal stacking. On the left are reported the calculations carried out with QE, while on the center and the the right the calculations made exploiting the CRYSTAL code. The former refers to the TVZP+ basis set, while the latter to the standard TVZP basis set.}
\label{fig:qe_vs_cry}
\end{figure*}

The choice of a suitable basis set (BS) is a critical approximation for the description of very low energy electronic structure of multilayer graphene. As mentioned above, we used atom-centered Hermite-Gaussian functions, which mimic atomic orbitals. In general, this kind of BSs require a lower number of basis set functions to provide good results and allow efficient computation of exact exchange integrals, but the quality of the approximation is more difficult to control with respect to plane wave BSs, and may be material-dependent. This is especially true for (semi)metallic solids, for which fewer basis sets have been devised and tested.
\\
Among all-electron Gaussian BSs optimized from solid state simulations, a popular choice which normally results in good accuracy is to use the POB-TZVP BSs \cite{POBbasis} which some of us have recently used for the study of ABC-stacked graphene multilayers.\cite{pamuk2017magnetic,RMGhenck,RMG_HM} 
%Generally speaking,  expanding an unknown function within the LCAO (linear combination of atomic orbitals) approximation in a set of known functions is not an approximation if and only if the basis set is complete.
%Some recommended basis sets are available at the CRYSTAL code homepage\cite{BasisSet}. Among such basis sets, for the C atoms we can find the C\_pob\_TZVP\_2012 one, that is the more extended. 
In order to validate its reliability for the present study, we have compared the band structures of the graphene multi-layers at the PBE level of theory with those obtained by means the plane wave calculations. As already mentioned, the convergence of the latter method with BS size is easier to verify, and will therefore be taken as reference. The results are reported in Fig. \ref{fig:qe_vs_cry}.
As it can be seen, the POB-TZVP basis set does not exactly recover the results of the plane waves for Bernal multilayers. For instance, if we consider the odd-numbered layers systems, the two Dirac cones are located in an erroneous position. The bottom Dirac cone is always below the Fermi level, and in some cases it drops down to energy lower than -30 meV, in this way it results absent in the energy region used in Fig. (\ref{fig:qe_vs_cry}) (see 5L and 7L). As a matter of fact, this Gaussian basis set get the populations of both electrons and holes wrong. The band dispersion for the even-numbered layers systems is wrong as well. Thus we have developed a new basis set for the C atom. We started from the def2-TZVP BS,\cite{DEF2basis} that has three more Gaussian functions in comparison to the POB-TZVP one: one \textit{s}, one \textit{d} and one \textit{f}. It can be freely downloaded from the basis set exchange site\cite{BasisSetExchange}. Since this basis set is devised for molecules, we have optimized the  exponents ($\alpha$) of the most diffused Gaussian functions, namely those with $\alpha < 0.3$, minimizing the energy. The resulting basis set is named TZVP+\cite{BSopt} for simplicity. The computed bands structure for TZVP+ is reported in Fig. \ref{fig:qe_vs_cry}. It can be seen that, in this case, the bands obtained by means of the Gaussian basis set closely resemble those computed with the plane waves: all of the main features of the plane waves bands are recovered. TZVP+, definitely, describes in a satisfactory way the interactions that are present in these systems.

\begin{figure*}[!h]
%\begin{minipage}[c]{0.2\linewidth}
\includegraphics[width=\textwidth]%angle=270]
{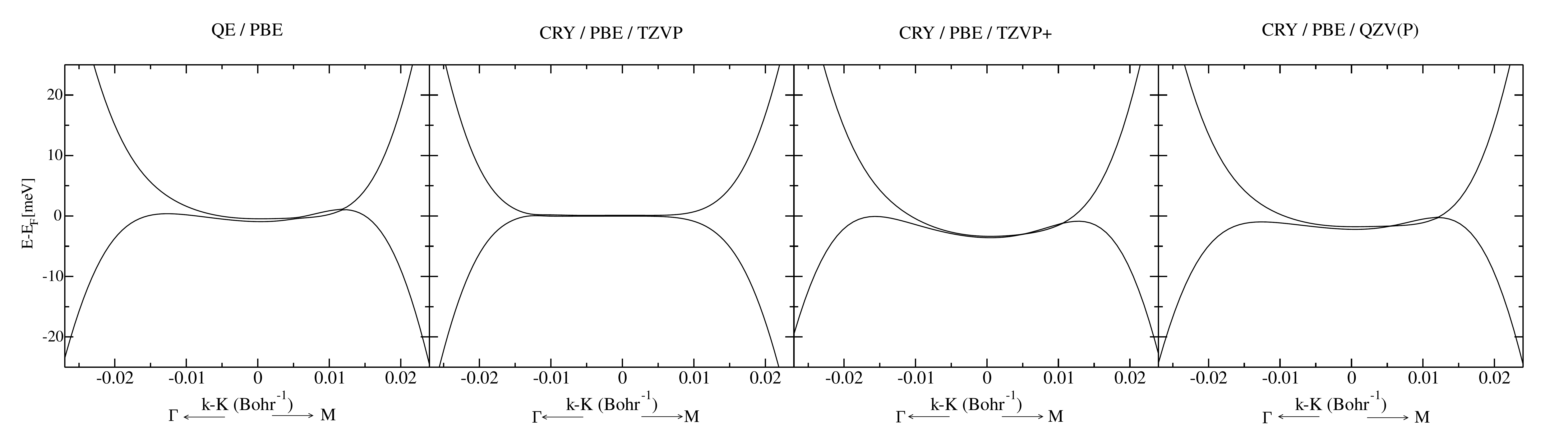} \smallskip \\
\includegraphics[width=\textwidth]%angle=270]
{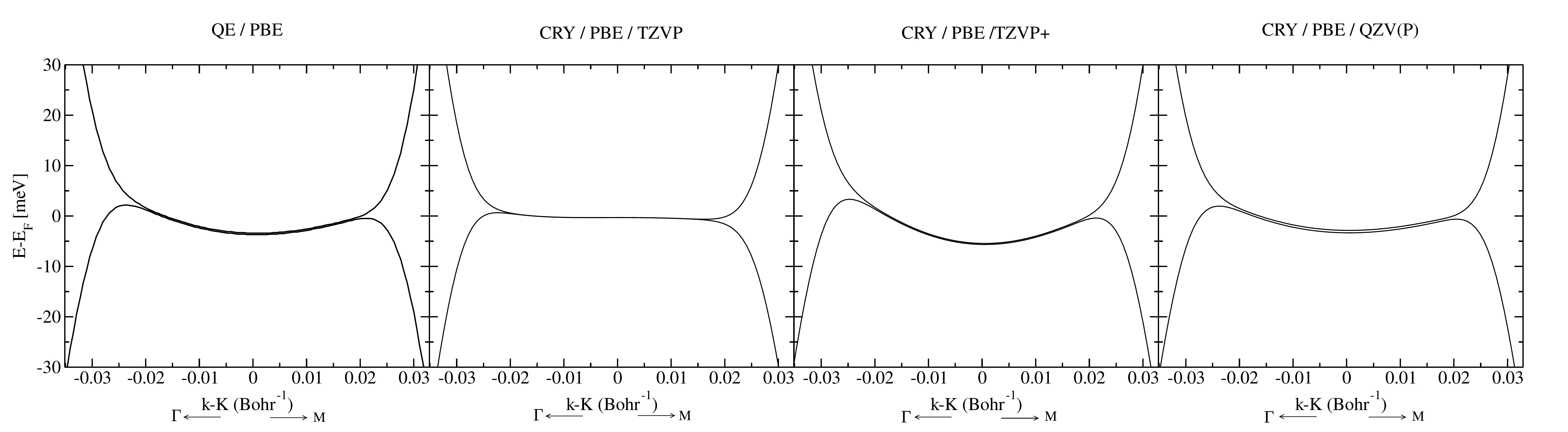}
\caption{PBE band structures for rhombohedral stacked multilayers (top: 6 layers, bottom: 14 layers) obtained with different approaches. On the left are reported the calculation carried out with QE, while the other calculations are made using the CRYSTAL code (the basis set specified in figure, of increasing size from left to right).}
\label{fig:qe_vs_cry_RMG}
\end{figure*}

In contrast, in the case of ABC-stacked multilayers  the POB-TZVP basis produces more satisfactory results, as there are no errors in the band occupations which appear for the AB stacking. Fig. \ref{fig:qe_vs_cry_RMG} shows PBE band structures for 6-layer and 14-layer). However, the curvature of the nearly-flat band at the Fermi level is underestimated by this Gaussian BS. The curvature, and thus the BS error, increases with the number of layers. Conversely, the TZVP+ basis overestimates the curvature by roughly the same amount. In order to obtain full agreement with the plane-wave calculations, we have to use a larger QZV(P) basis, which we have also reoptimized starting from a def2-QZVP molecular BS, but where we have reduced the high angular momentum polarization functions to the ones of the TZVP+ basis.

 We report here also the PBE0 result for 14 layers, for which the larger discrepancies between BSs are observed. Using the most accurate QZV(P) basis, we also obtain an enhanced curvature of the nearly-flat band with respect to the POB-TZVP results, with a bandwith of 8 meV as opposed to less than 2 meV reported in previous publications,\cite{RMGhenck} which are however unaffected in the qualitative results and conclusions.

\begin{figure}[!h]
%\begin{minipage}[c]{0.2\linewidth}
\centering
\includegraphics[width=0.45\textwidth]%angle=270]
{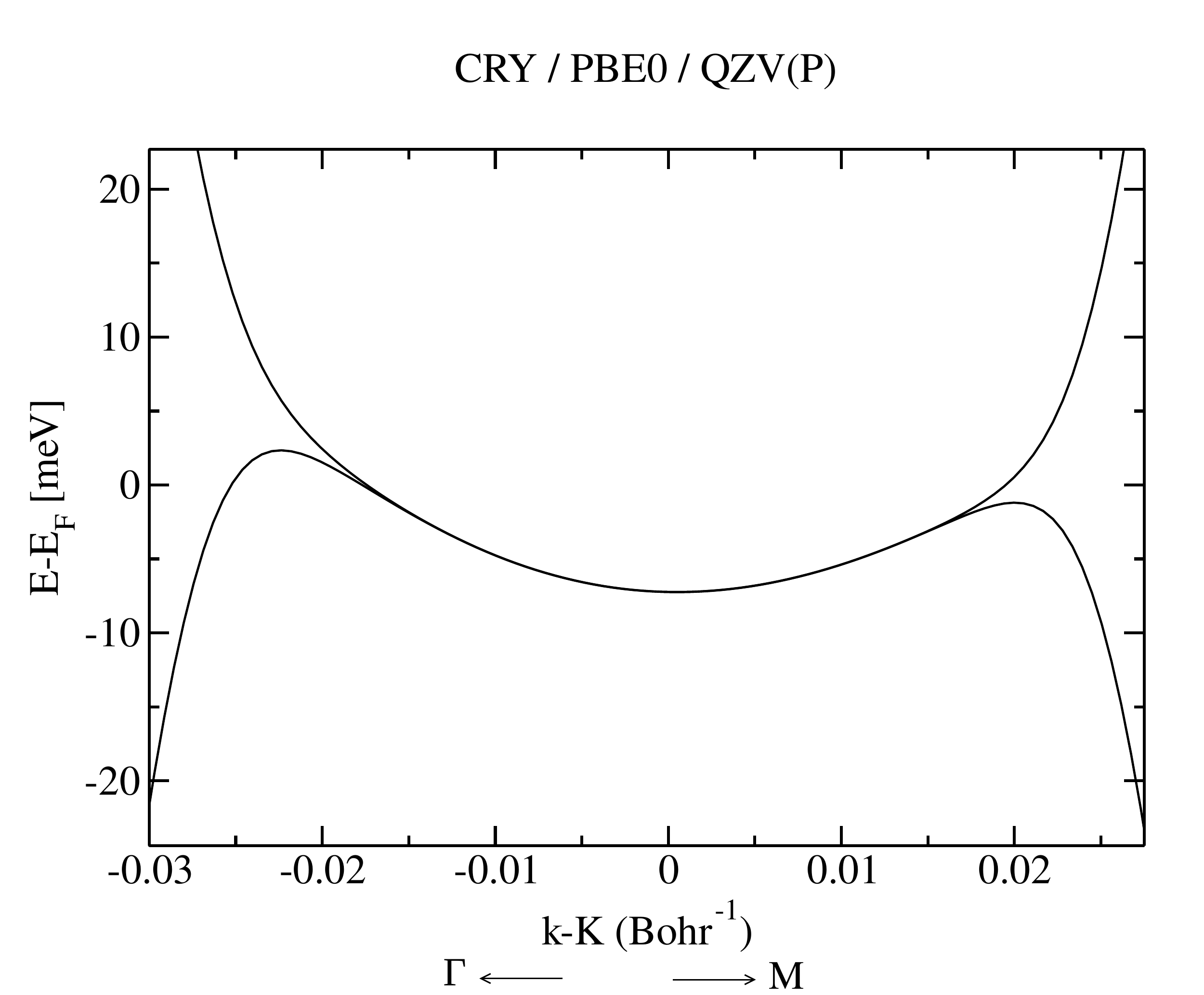}
\caption{PBE0 band structure for rhombohedral stacked 14-layer, obtained with the QZV(P) basis (the most accurate basis set size).}
\label{fig:PBE0_RMG}
\end{figure}

\section{Results for Bernal multilayers\label{sec:results}}

\subsection{Bands Decomposition}

\subsubsection{Structure of multilayer Bernal graphene\label{sec:3Lvs4L_PBE}}

In Fig. \ref{fig:bandepbe0} (top panels) we report the crystal structures and atom labeling, for
$N=3,4$ multilayer graphene calculated using both the PBE and PBE0 functional. 
It is important to note that in the case of Bernal stacking, the two inequivalent atoms
per layer have different connectivity to out-of-plane nearest neghbours. One of the two
atoms in each layer is vertically connected to two out-of-plane nearest neighbour atoms
in nearby bilayers, i.e. it has one neighbour exactly on top and one exactly below. We label this atom to be of the A-type. The other no out of plane neighbours in nearest layers, neither on top nor below, and we label it as being of the B-type\cite{mcclure1957band,slonczewski1958band}. A and B type atoms alternate within each layer, however A atoms lay always exactly on-top or below of A atoms in other neighbouring layers. The out-of-plane hopping between nearest neighbours A-atoms is normally labeled $\gamma_1$.

\subsubsection{Trilayer Graphene}

The PBE electronic structure of trilayer graphene close to the Fermi level is composed of three bands: a gapped and slightly hole-doped Dirac cone and two gapped massive bands,
as shown in Fig. \ref{fig:bandepbe0}. The band-overlap between the hole doped part of the gapped Dirac cone and the empty 
massive band is approximately $1.8$ meV.
The bonding part of the gapped Dirac cone is formed mainly by 2p$_z$ orbitals 
of one of the two carbon atoms in the two outermost layers, namely atoms $2$ and $6$. These atoms are of the A-type.
The non-bonding part of the gapped Dirac cone as well as the empty massive bands are, on the contrary, mostly formed by 2p$_z$ orbitals
of B-type carbon atoms in the outermost layers, atoms $1$ and $5$. Finally, the occupied massive band is mostly formed by atom $4$ in the innermost layer, that is
of B-type. Atom $3$ (A-type) does not contribute to the electronic structure
in this energy region. Thus, all massive bands are mainly due to B-atoms (i.e. atoms having no neighbouring
atoms exactly on top and thus not connected by the $\gamma_1$ hopping).\newline
The electronic spectrum is not particle hole symmetric, mainly because the gapped Dirac cone is shifted with respect to the center of the massive bands. However, even the massive bands themselves are not particle-hole symmetric one with respect to the other.
In a mimimal tight-binding model in which only the $\gamma_0$ (in-plane nearest-neighbours hopping)
and $\gamma_1$ (1$^{st}$ out of plane neighbor) are considered, the spectrum is completely particle hole symmetric. As additional in-plane hoppings do not shift in energy the Dirac
cone but, simply change the slope of the Dirac bands, it follows that the position of the Dirac point with respect to the massive bands can only be determined
by long-distance out-of-plane hoppings. Previous work \cite{Koshino_Exchange} suggested that this could be due to the $\gamma_2$ hopping,
namely the hopping between atom $1$ and $5$,  i.e. the vertical hopping process between two B-type atoms laying in next nearest neighbour layers. This hopping is, of course, relevant only for $N\ge 3$. 
Moreover, we can also note that, even if the trilayer is not gapped in PBE, the large Fermi velocity of the Dirac cone leads to a small density of states at the Fermi level
and the appearing of a pseudogap of $\approx 3.5$ meV, as it can be seen in Fig. \ref{fig:bande_vs_doss_dispari}.\newline

PBE0 exibits a similar electronic structure, as a matter of fact we can retrieve the gapped Dirac cone and the two massive bands as well. However in this case, the exact exchange in the exchange-correlation energy, has slightly changed the atomic contributions to the electronic bands. As in the previous case, the main character of the massive bands comes from B-type. However, for what concerns the Dirac cone,
now the bonding part is due to atom $1$ and $5$ (B type) while the non-bonding part is due to
atoms $2$ and $6$ (A type), the opposite of the PBE case. The Dirac cone are moved to higher energies, as compared to PBE, and the gap between them decreases from 10 meV (PBE) to 8 meV (PBE0). 

It is worthwhile to stress that our PBE0 calculation is in excellent agreement with the GW calculation of Ref. \onlinecite{PhysRevB.89.035431}.

\subsubsection{Four layer graphene}

The low-energy electronic structure of four layer graphene is composed of four massive bands. There is a small band overlap between the valence and conduction bands so that the system is metallic in PBE, as shown in Fig. \ref{fig:bandepbe0} and \ref{fig:bande_vs_doss_pari}. All bands in a 30 meV energy
window from the Fermi level are formed by the 2p$_z$ states of B atoms 
(i.e. not connected by $\gamma_1$). 
More specifically, the bands are formed by the 2p$_z$ states of atoms $1$ and $8$ in the outer layers, 
while the occupied bands are formed by atoms $4$ and $5$, namely atoms in the inner layers. This is very similar to the three-layer case, namely
the character of the massive bands is due to atoms not involved in $\gamma_1$ hopping. Interestingly, there is an exchange in the band character close
to the Fermi level at the top of the lowest energy bands along MK and at the bottom of the two highest energy bands.
This character exchange is present also in a minimal model including only $\gamma_0$ and $\gamma_1$\cite{Koshino_Exchange},
albeit the form of the bands in the PBE calculation, including all possible hopping processes and metallic screening, is substantially more complex.
Unlike the three-layer system, in this case the introduction of the exact exchange doesn't change the composition of the bands. The only relevant change is the missing crossing bands along the MK high-symmetry direction.   

\subsection{Electronic structure of thicker flakes\label{sec:PBEvsPBE0}}

\subsubsection{Odd number of layers}

\begin{figure*}
%\begin{minipage}[c]{0.2\linewidth}
%\begin{center}
%\hspace{10.0truecm}
\includegraphics[scale=0.2]%angle=270] 
{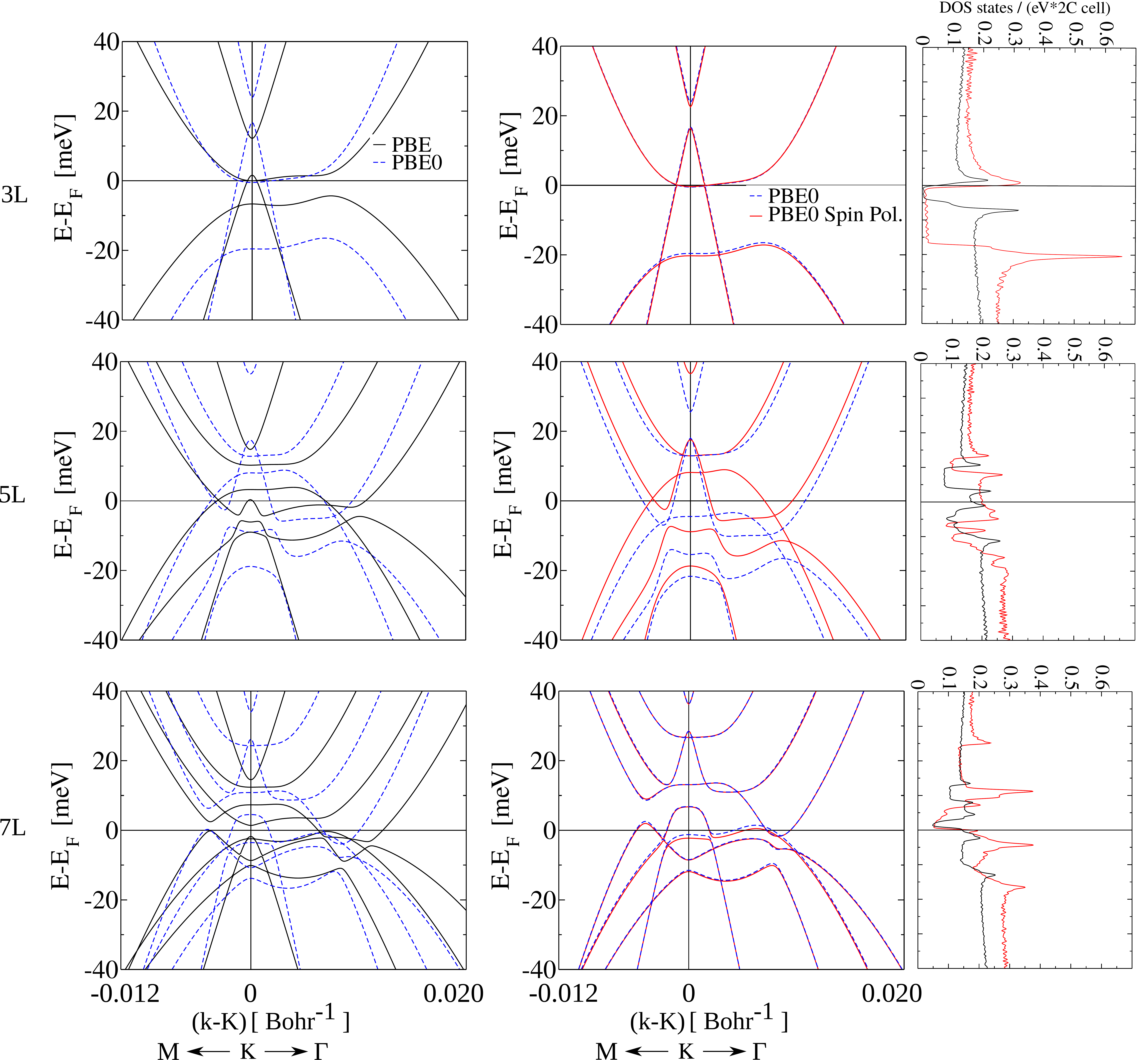}
%\end{minipage}
\caption{Left panels : PBE versus PBE0 electronic structure for Bernal graphene multilayers with $N=3,5,7$. Middle panels: PBE0 versus PBE0 spin polarized electronic structures. The PBE0 magnetic and PBE non-magnetic density of states are reported in the rightmost panels.}
\label{fig:bande_vs_doss_dispari}
%\end{center}
\end{figure*}

The PBE and PBE0 electronic structures for $N=3,5,7$ are shown on the left panels of Figure \ref{fig:bande_vs_doss_dispari}. We note that all structures are metallic, both in PBE and PBE0. In all cases, in a $40$ meV window from the Fermi level there are: (i) a gapped Dirac cone and (ii) $(N-1)$ parabolic-like bands. The latter are related to the massive bands
in graphite and the separation between them is related to the k$_z$ dispersion of the electronic structure in graphite. The overlap between the Dirac band and the massive bands depends
sensibly on the number of layers as well as on the exchange correlation functional used
in the calculation. Both at the PBE and PBE0 level, the three and seven layers case are pseudogapped close to the Fermi level (directly below or above), the pseudogap being larger for the PBE0. The pseudogap is larger for the $N=3$ than for $N=7$,
in qualitative disagreement with the gap inferred from transport
data in Ref. \onlinecite{nam2018family}.

The action of the exchange interaction, despite differing in details for
different $N$, has some features common to all flakes composed of odd $N$. 
The first important point is that the slope of the
Dirac cone close to K as well as the slopes of the massive bands far from K
are increased by the exchange interaction. This corresponds to a well known
\cite{Ohta2006,Zhou2006,Sugawara2007} renormalization of the in-plane hopping matrix elements that is
present both in graphene and graphite. Second, the exchange interaction tends to open gaps between bonding and non-bonding massive bands with opposite concavities. These two effects, combined with the conservation of the
number of electrons, results in an upshift of the Dirac cone with respect to the massive
bands. It is crucial to remark that this is exactly the opposite of the effect
predicted in Ref. \onlinecite{Koshino_Exchange} based on self-consistent tight binding
calculations including empirical exchange interaction. The main reason is that
in Ref. \onlinecite{Koshino_Exchange} the exchange interaction simply renormalizes $\gamma_2$, but not $\gamma_1$ that is kept at the experimental value in disgragreement with what has been shown to occurr in graphene.
Including part of the renormalization in the
$\gamma_1$ and varying the effect of the interaction on $\gamma_2$ only
introduces an error and is equivalent to an {\it ad hoc} tuning of the position of the 
Dirac band. This underlines the need of performing explicit calculations of the exchange interaction beyond the tight binding approach.

\subsubsection{Even number of layers}

In the case of an even number of layers, the electronic structure in a $40$ meV window from the Fermi level is composed of
$N$ massive bands. As in the odd number of layer case, the exchange interaction
tends to open gaps between the bonding and antibonding massive bands. However, here, only the four layer becomes completely gapped (the gap being $3.2$ meV) and insulating in PBE0
due to the removal of the weak band-overlap present at the PBE level. For 
$N=6,8$, the flakes are metallic both in PBE and PBE0, however they develop small pseudogaps at the PBE0 level ($1.8$ meV and $1.6$ meV for 6L and 8L respectively). Even in this case, the magnitude and the behaviour with thickness of the
(pseudo) gaps disagree with experimental data given in Ref. \onlinecite{nam2018family}.   
\begin{figure*}
%\begin{minipage}[c]{0.2\linewidth}
%\begin{center}
%\hspace{10.0truecm}
\includegraphics[scale=0.2]%angle=270] 
{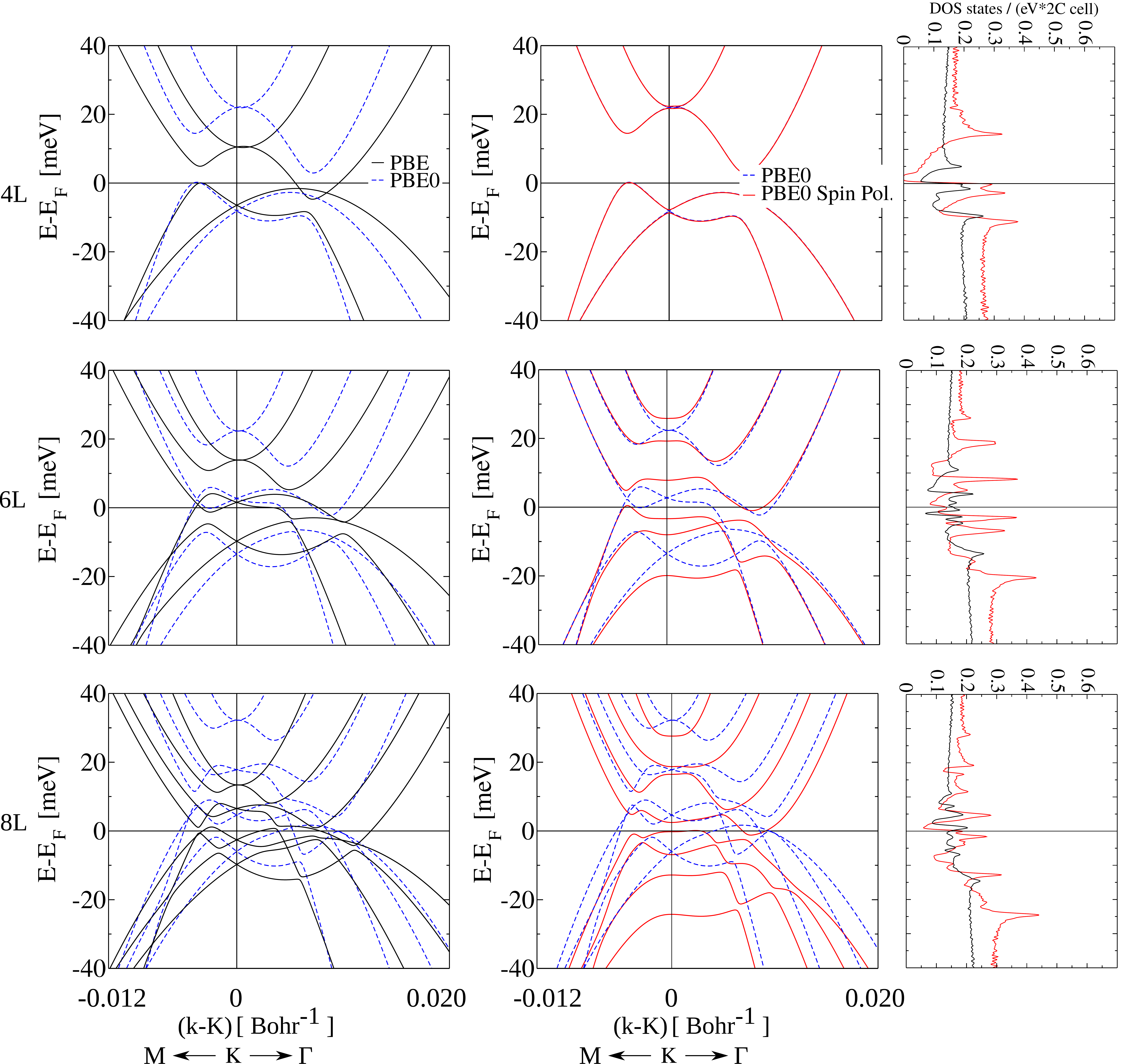}
%\end{minipage}
\caption{Left panels : PBE versus PBE0 electronic structure for Bernal graphene multilayers with $N=4,6,8$. Middle panels: PBE0 non spin-polarized versus PBE0 spin-polarized electronic structures. The PBE0 magnetic and PBE non-magnetic density of states are reported in the rightmost panels.}
\label{fig:bande_vs_doss_pari}
%\end{center}
\end{figure*}

\subsection{Magnetic states\label{sec:magnetism}}

\begin{figure*}[!h]
%\begin{minipage}[c]{0.2\linewidth}
\includegraphics[scale=0.25]%angle=270] 
{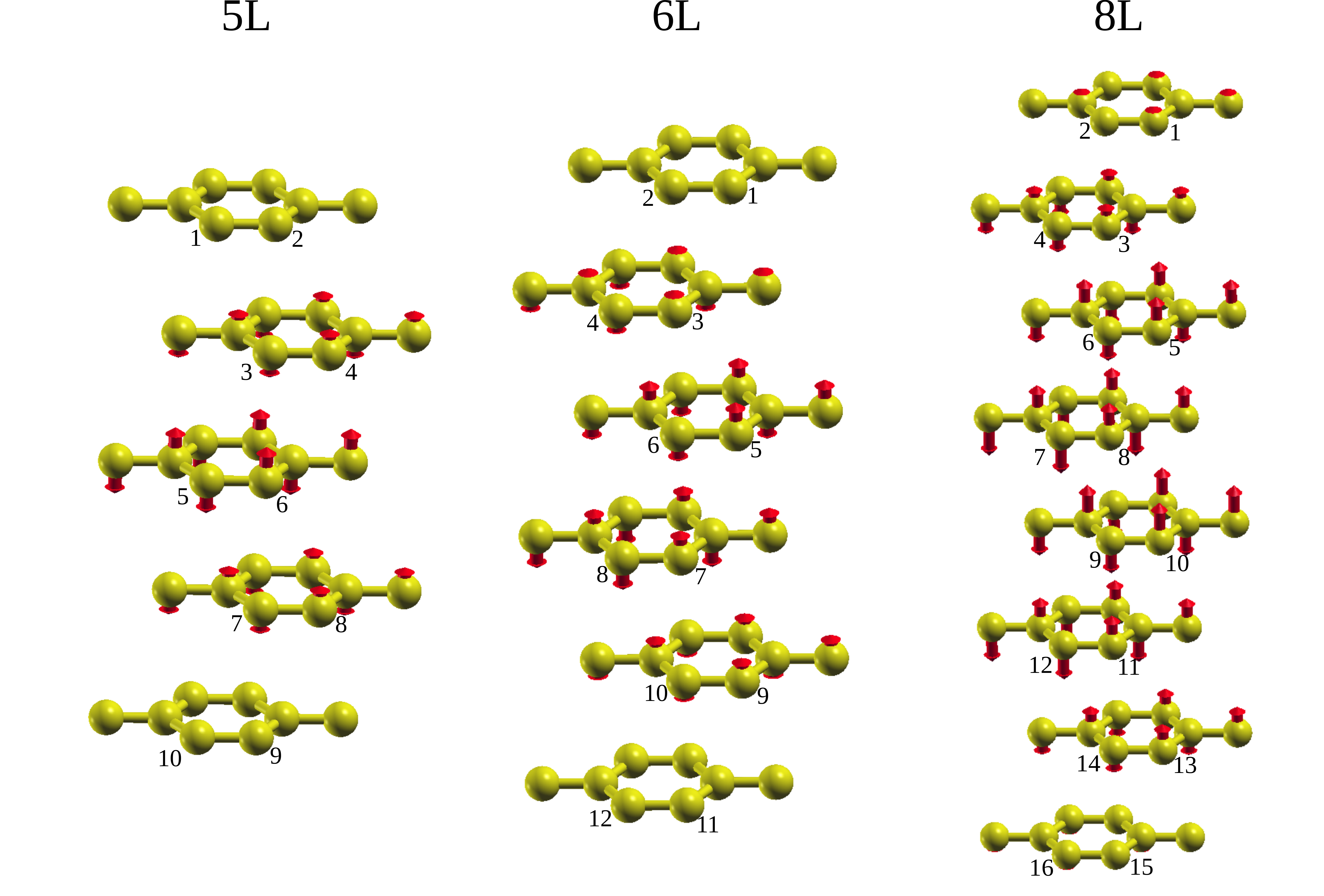}
%\end{minipage}
\caption{Schematic representation of the magnetic ground state for 5L, 6L and 8L systems respectively for Bernal graphene multilayers. The spins (as arrows) and atom numbering are reported.}
\label{fig:spinsketch}
\end{figure*}

As several other carbon base systems such as rhombohedral stacked multilayer graphene \cite{Henni_NL,pamuk2017magnetic,RMGhenck,Lautrilayer,Lauquadrilayer},
twisted bilayer graphene\cite{cao2018unconventional,cao2018correlated, GuineaPhysRevLett.119.107201} and diamond(111)\cite{Betul_Diamond}  have been suggested to
host magnetic states, it is meaningful to verify if magnetism can be stabilized in this system.
For this reason we run spin polarized  calculations in PBE and PBE0. It is worthwhile to notice that in order to stabilize magnetism
an ultradense grid of k-points needs to be used. Indeed, if the region close to the point K and in a $20$ meV window from the Fermi level is not correctly
sampled, the solution will always be non-magnetic. Within PBE we never managed to stabilize a magnetic state.
On the contrary, in PBE0  we stabilize a globally antiferromagnetic state.  
The starting magnetic state of our simulation is antiferromagnetic both globally and within each layer with anti-ferromagnetic coupling between layers. 
Only few flakes stabilize sizeable magnetic moments, namely the 5L, 6L and 8L, all the other flakes had a negligible atomic magnetic moment ($\leq \times 10^{-4} \mu_B$). All the magnetic ground
states have three common features: first, all the magnetic states are globally antiferromagnetic (the bands are twofold degenerate in spin), second, each layer is ferrimagnetic, 
namely two close atoms have opposite spin but with slight different magnitude of the magnetic moments and, finally, the magnitude of the spin drops significantly in going
from the inner layers to the outer ones. The magnetic states are depicted in figure \ref{fig:spinsketch} and the magnetic moments reported in table \ref{tbl:spinmagnitude}.
These results are consistent with a previous investigation performed on rhombohedral stacked graphene\cite{pamuk2017magnetic}: namely the presence of stable magnetic states in multi-layer graphene systems. However, here there are two main differences: (i) The magnetic moments increase going from the outer layers to the center (while in ABC graphene multilayers it is the opposite) and (ii) the magnitude of the magnetic moments is much smaller in the present case.  The main reason for this difference holds in the nature of the character forming the massive bands. While
in ABC graphene multilayers the massive bands are mostly formed by atoms in the outermost layers, in Bernal graphene multilayer the massive bands are formed by B atom  types and the bonding
ones mostly by B atoms in the innermost layers.
\begin{table}
  \caption{The magnitude of the spin of each atom in units of $10^{-3}$ $\mu_B$ are reported. Due to the symmetry of the magnetic states established, we report for 5L spin up to $\mu_{6}$, since we have $\mu_{1}=\mu_{10}$, $\mu_{2}=\mu_{9}$, $\mu_{3}=\mu_{7}$ and $\mu_{4}=\mu_{8}$. For the systems with an even number of layers (6L and 8L) we found the following symmetry : $\mu_i = -\mu_{2N−i+1}$ for $i=1,N$. Hence, we report the spin up to $\mu_N$.}
  \label{tbl:spinmagnitude}
  \begin{center}
  \begin{tabular}{lcccccccc}
    \hline
    \hline
$N$  &  $\mu_1$ &   $\mu_2$  &  $\mu_3$  &   $\mu_4$ & $\mu_5$ & $\mu_6$ & $\mu_7$ & $\mu_8$\\
\hline
5      &  1.56  &   -1.48    &  2.10     &  -2.29    &  2.40  & -2.22  \\

6      &  0.71  &    -0.65   &  1.08     &  -1.26    &  1.70  &  -1.47    \\

8      &  0.93  &    -0.83   &  1.44     &  -1.67    &  2.20  &  -1.90  &  2.21  &  -2.24   \\
    \hline
    \hline
  \end{tabular}
    \end{center}
\end{table}
Magnetism has some effects on the electronic structure of 5, 6 and 8 layers, as shown in Figs. \ref{fig:bande_vs_doss_dispari} and \ref{fig:bande_vs_doss_pari}, mainly opening gaps and removing
exact crossings at the $K$ high-symmetry point. However, not enough to make the system insulating. It slightly increases the pseudogaps found at the non magnetic PBE0 level.  

%\section{Comparison with experimental data}
%\label{exp_vs_theory}

%{\bf to be finished}.
\section{Results for Rhombohedral multilayers\label{sec:results}}

In our previous works \cite{pamuk2017magnetic,RMG_HM, RMGhenck} carried out with the
POB-TZVP basis set, we found that multilayer graphene with rhombohedral stacking has
a magnetic ground state (at least for thicknesses up to 14 layers) with an ultraflat
surface state. Given the lower accuracy of POB-TZVP basis sets in the 100 meV energy region
with respect to those developed in the current work, it is worthwhile to re-evaluate 
electronic structure and magnetic properties in this energy region using the more accurate
QZV(P). This is the purpose of the current section. We consider $6$ and $14$ layers. Calculations
on three layers were recently carried out by us in Ref. \onlinecite{LauCalandra} using the QZV(P)
finding small differences with the case of Ref. \onlinecite{pamuk2017magnetic,RMG_HM}. However,
the situation could change in thicker samples as the main difference between the QZV(P)
and POB-TZVP basis sets is the treatment of long distance hoppings.

\subsection{6 layers}

\begin{figure}[!h]
%\begin{minipage}[c]{0.2\linewidth}
\includegraphics[scale=0.3]%angle=270] 
{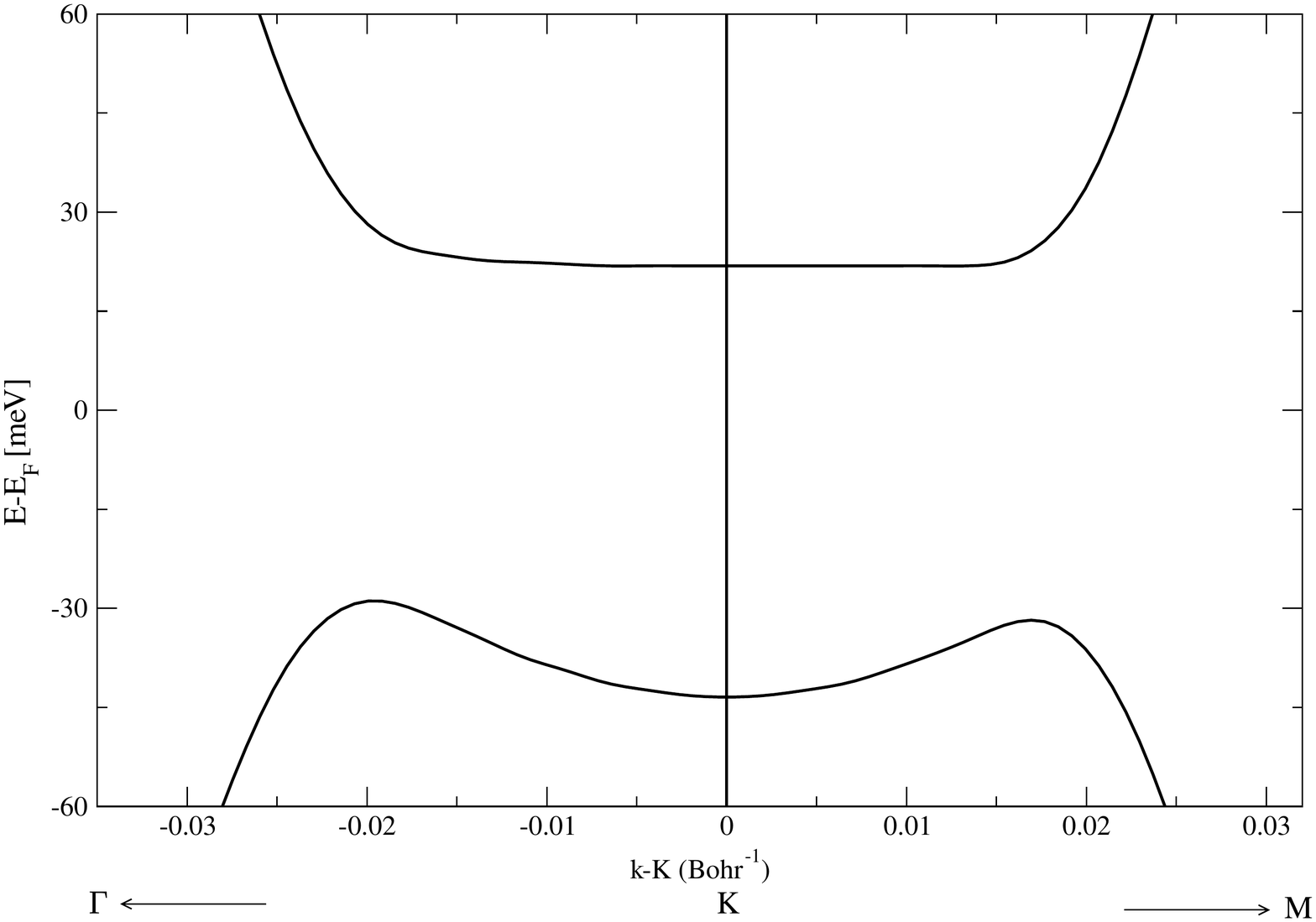}
%\end{minipage}
\caption{Magnetic PBE0 electronic structure of 6 layers ABC graphene calculated using the most accurate QZV(P) basis set.}
\label{fig:6L_PBE0}
\end{figure}

Within PBE0, the ground state of 6 layers ABC graphene is antiferromagnetic. Qualitatively and quantitatively,
practically no differences are found for what concerns the magnetic state with respect to Ref. \onlinecite{pamuk2017magnetic}.
The ground state is the layer antiferromagnet, usually labeled LAF \cite{Lautrilayer}, with the largest magnetic moments
concentrated in the outermost layers.
The electronic structure, however, does show some differences. The indirect gap is of similar magnitude ($\approx 50$ meV), 
but the dispersion and effective mass of the valence and conduction bands depends substantially on the basis set used. In particular, by using the most
accurate QZV(P), the dispersion of the top
of the valence band from its maximum along $\Gamma K$ to the K-point is approximately 13 meV , while the
less accurate POB-TZVP basis set leads to 7-8 meV dispersion, namely $\approx 35-40\%$ underestimation.
On the contrary, the curvature of the bottom of the conduction band at K is substantially smaller by using the more accurate
QZV(P) basis set (the conduction band is almost flat ar K) with respect to the POB-TZVP basis set. 

\subsection{14 layers and comparison with ARPES experiments}

\begin{figure}[!h]
%\begin{minipage}[c]{0.2\linewidth}
\includegraphics[scale=0.3]%angle=270] 
{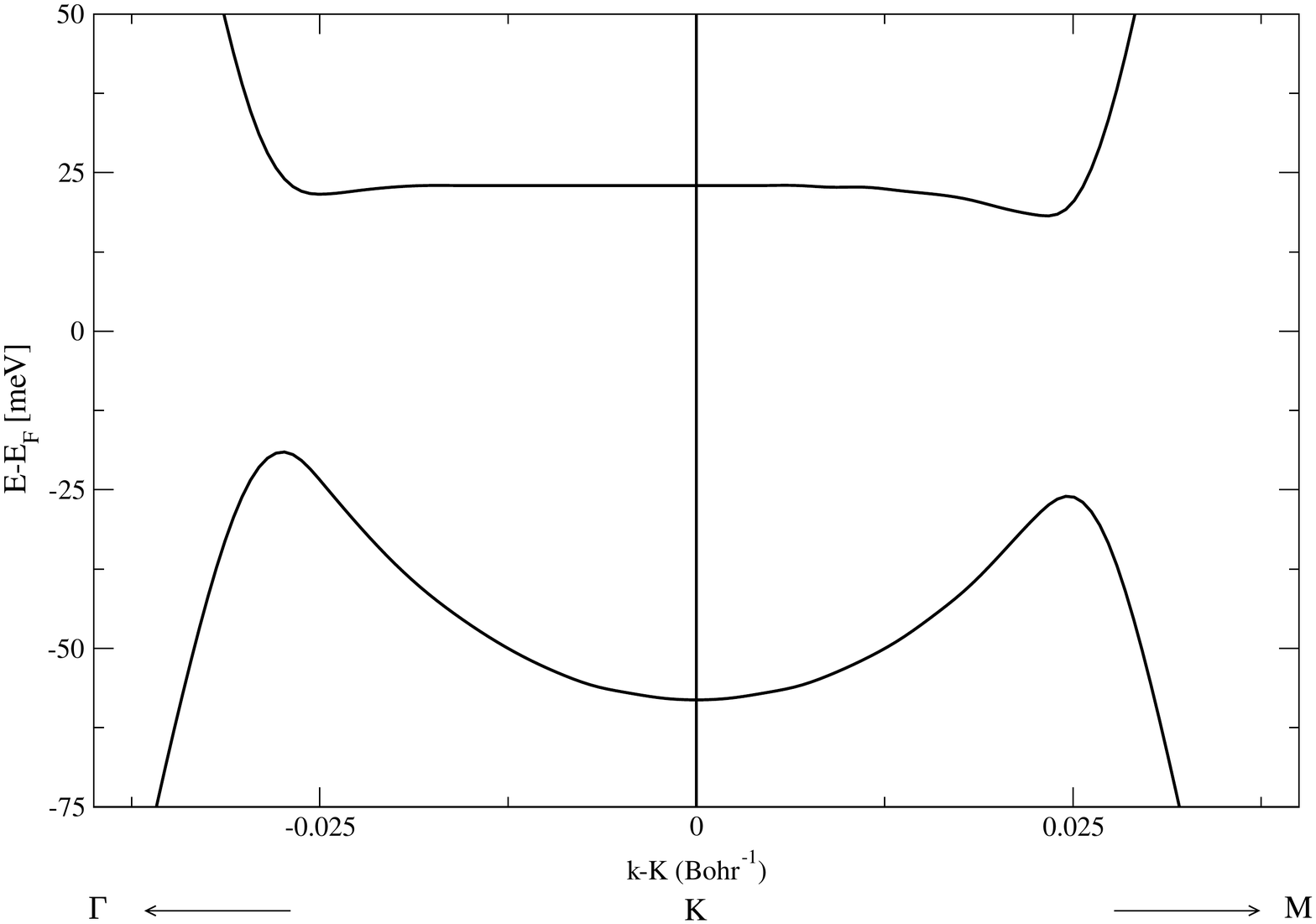}
%\end{minipage}
\caption{Magnetic PBE0 electronic structure of 14 layers ABC graphene calculated using the most accurate QZV(P) basis set. }
\label{fig:14LPBE0}
\end{figure}

\begin{figure}[!h]
%\begin{minipage}[c]{0.2\linewidth}
\includegraphics[scale=0.3]%angle=270] 
{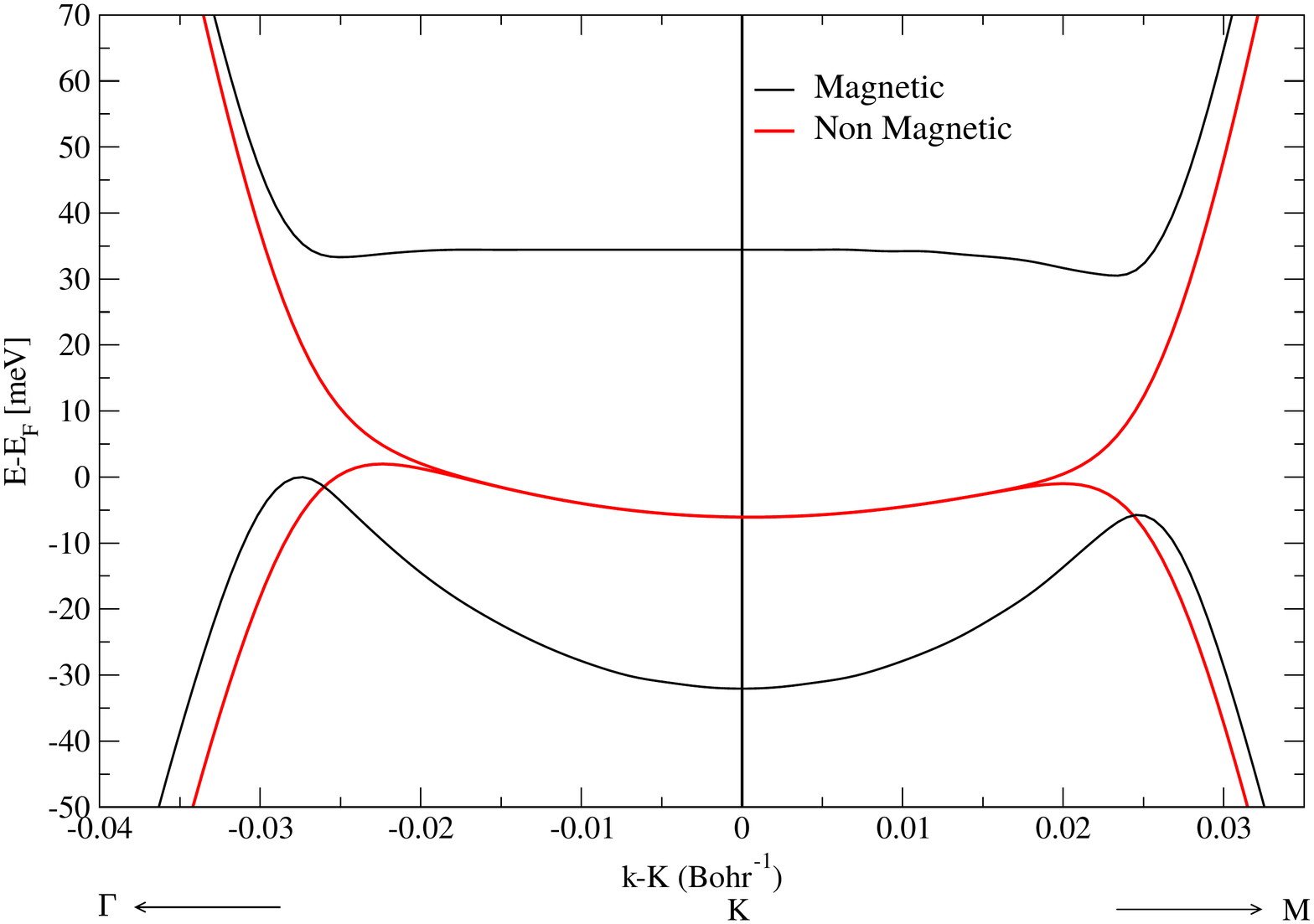}
%\end{minipage}
\caption{Magnetic and non-magnetic PBE0 electronic structure of 14 layers rhombohedral stacked multilayer graphene calculated using the most accurate QZV(P) basis set. In order to directly compare with ARPES in Ref. \onlinecite{RMGhenck}, we reduce Fermi velocity of 16\%. Indeed, it is well known that PBE0 gives a too large exchange renormalization of the Fermi velocity with respect to GW in graphene\cite{Zhou2006}.
To better compare the two calculations the Fermi level for the magnetic insumating cas has been set at the  valence band top.}
\label{fig:14LPBE0resc}
\end{figure}

The 14 layers calculation is important as the dispersion and curvature of the valence band close to the point K have been used in ARPES \cite{RMGhenck} to identify the occurrence
of a possible magnetic state. In experiments,  the band dispersion of the valence band top, from its minimum at K to the maximum along $\Gamma K$ was estimated to be $\approx 25$ meV. As shown in Sec. \ref{sec:QZVP+}, within the PBE0 approximation, in the non-magnetic case and using the  POB-TZVP basis set, the dispersion is smaller than 2 meV, but it increases to $\approx 8$ meV when the QZV(P) basis set is used. In the magnetic case, the band dispersion using the most accurate QZV(P) basis set is shown in Fig. \ref{fig:14LPBE0} and is approximately 40 meV. This demonstrates
that the curvature in the magnetic case is substantially enhanced with respect to the non-magnetic case, in agreement with the results of Ref.  \onlinecite{RMGhenck}. In order to compare with experiments, some care is needed, as it is well knon that
for graphene the PBE0 exchange and correlation functional overestimates the Fermi velocity of approximately 16\% \cite{Zhou2006}.
By applying this reduction, the dispersion of the top of the valence band is $\approx 33$ meV,   not too far from the 25 meV estimated in 
ARPES experiments \cite{RMGhenck}.
The difference between the magnetic and non-magnetic band structure is substantial, as
shown in Fig. \ref{fig:14LPBE0resc}, and the ARPES results clearly show a better agreement with a magnetic electronic structure, as concluded in Ref. \onlinecite{RMGhenck}.

\section{Conclusions}
In this paper, we study the electronic structure of multilayer graphene by means of hybrid functionals within a localised basis set approach. We show that basis sets normally considered as very accurate, such as the  POB-TZVP one, fail substantially 
in predicting the electronic structure of multilayer graphene with Bernal stacking. For rhombohedral stacking the error is much smaller, as the gap and magnetic state are essentially identical to those calculated with more extended basis sets, but the curvature and dispersion of the valence band top is substantially underestimated (the effect is negligible for trilayers but become important for thicker multilayers).
We solved this problem by developing two new basis sets that perfectly reproduce the plane-waves calculations and lead to a very accurate
description of the electronic structure in the 100 meV range from the fermi level.

With these new basis sets, we study the effects of exact exchange as included in the PBE0 functional on the electronic structure.
For Bernal multilayers, we found that in all case (except 5 layers) the exchange interaction opens gaps (4 layers) or pseudogaps (3,6,7,8 layers). However, both the size and the thickness dependence of the pseudogaps disagree with experiments\cite{nam2018family}. For 5, 6 and 8 layers, the ground state is found to be magnetic with very small magnetic moments of the order of $10^{-3}\mu_{B}$. These magnetic moments have very small effects on the non-magnetic electronic structure.
 The magnetic state is such that the inner layers are weakly ferrimagnetic in the graphene plane, the surface layers have vanishing momenta and the global state results to be anti-ferromagnetic. Thus, unlike the ABC multilayer graphene, the magnitude of the magnetic moments increases going from the outer layers to the center. 
Our results suggest that the insulating nature of AB-stacked multilayer graphene can be due to two phenomena. One possibility is that many body effects beyond the mean field single-particle theory are responsible for the gap opening. However, given that our results for ABC trilayer graphene are practically indistinguishable 
from non-magnetic GW calculations in Ref. \onlinecite{PhysRevB.89.035431}, it is not obvious how to improve this result. A second possibility is that the gap is not intrinsic and is triggered by some kind of external interaction (small electric fields, residual doping, asymmetry in the samples along the z-axis...). More work is needed to clarify this issue.

Finally, we revaluate the effect of more accurate basis sets for the case of 6 and 14 graphene multilayer with rhombohedral stacking.
While the magnetic ground state properties and the gap are essentially unchanged with respect to previous calculations using 
less accurate basis sets \cite{pamuk2017magnetic,RMGhenck,RMG_HM}, the dispersion of the valence band top (conduction band bottom) was underestimated (overestimated). The underestimation of the valence band top energy dispersion is larger in the non-magnetic case, while it is smaller in the magnetic case. At odds with the Bernal case, in the rhombohedral multilayers the stabilization of larger magnetic moments and the favourable comparison with ARPES suggest that mean field theory including the exchange interaction correctly
describes the ground state properties of the system.

 Our calculations represent a new benchmarck for the theoretical description
of  the low energy physics graphene multilayers beyond standard density functional theory with semilocal kernels and will be a reference for future manybody calculations.

\section{Acknowledgements}
We acknowledge support from the Graphene Flagship
Core 2 grant number 785219 and Agence Nationale de la Recherche under 
references ANR-17-CE24-0030.
This work was performed using HPC resources from GENCI, TGCC,
CINES, IDRIS (Grant 2019-A0050901202 and The Grand Challenge Jean Zay).

\end{document}